\documentclass[letterpaper]{jpconf}
\usepackage{graphicx}

\def\chpt{\raise0.4ex\hbox{$\chi$}PT}
\def\schpt{S\raise0.4ex\hbox{$\chi$}PT}
\def\MeV{{\rm Me\!V}}
\def\GeV{{\rm Ge\!V}}

\vspace{-1.5cm}
\begin{document}
\title{Light hadrons in 2+1 flavor lattice QCD\footnote{To appear in
the proceedings of the First Meeting of the APS Topical Group on Hadronic
Physics, Fermilab, Batavia, Illinois, Oct. 24-26, 2004}}

\author{Urs M.~Heller (For the MILC Collaboration)}

\address{American Physical Society, One Research Road, P.O. Box 9000,
Ridge, NY 11961-9000, USA}

\ead{heller@aps.org}

\begin{abstract}
This talk will focus on recent results by the MILC collaboration
from simulations of light hadrons in 2+1 flavor lattice QCD.
We have achieved high precision results in the pseudoscalar sector,
including masses and decay constants, plus quark masses and
Gasser-Leutwyler parameters from well controlled chiral perturbation
theory fits to our data. We also show spectroscopy results for
vector mesons and baryons.
\end{abstract}

\section{Introduction}

The MILC Collaboration has been engaged in the study of QCD with three
flavors of improved staggered sea quarks~\cite{ASQTAD} for some time. Our
goals are to determine the hadron mass spectrum, the properties of light
pseudoscalar mesons, the topological susceptibility, the decay properties
of B and D mesons, and the behavior of strongly interacting matter at high
temperatures, all in as realistic simulations as possible. In this talk
I will concentrate on the light pseudoscalar sector and the light hadron
spectroscopy~\cite{PSEUDO,SPEC}.

While we have made simulations with lighter u and d quark masses
(with $m_u=m_d=m_l$) than have been reached before, to simulate at the
physical mass value has not been possible so far even with the improved
``asqtad'' quarks~\cite{ASQTAD}. We have created gauge field
ensembles at two lattice spacings, $a \sim 0.125$ and $\sim 0.09$ fm, with
the strange quark mass $m_s$ fixed near its physical value and several
light quark masses, as low as $m_l=0.1\, m_s$. This allows controlled
chiral extrapolations to the physical light quark mass, and from the two
lattice spacings, extrapolations to the continuum limit, $a \to 0$.  The
physical spatial size of our gauge field ensembles is $L \ge 2.5$ fm.
At one set of simulation parameters we created ensembles with two physical
sizes, $L = 2.5$ and 3.5 fm, to check for finite volume effects. These
were found to be small.

All gauge field ensembles created by the MILC collaboration are made
available to other researchers at the ``NERSC Gauge Connection''. They
have been used, for example, by the Fermilab, HPQCD and UKQCD
collaborations for the study of heavy quarkonia and heavy--light mesons. A
comparison of quantities that can be computed on the lattice with small
systematic uncertainties and that are well known experimentally shows
agreement within errors of 1--3\%~\cite{PRL}.

\section{The light pseudoscalar sector}

In the pseudoscalar sector we have the most accurate data with many
partially quenched measurements, {\it i.e.}, measurements of pseudoscalar
masses and decay constants with valence quark masses different from the
sea quark masses. Fits, even at a single lattice spacing, to continuum
partially quenched chiral perturbation theory (\chpt) do not
work~\cite{PSEUDO}. Lattice effects, in particular the ${\cal
O}(\alpha_s^2 a^2)$ taste breaking effects of the improved staggered
fermions, need to be taken into account.

\begin{figure}[h]
\begin{minipage}{18pc}
\includegraphics[width=18pc]{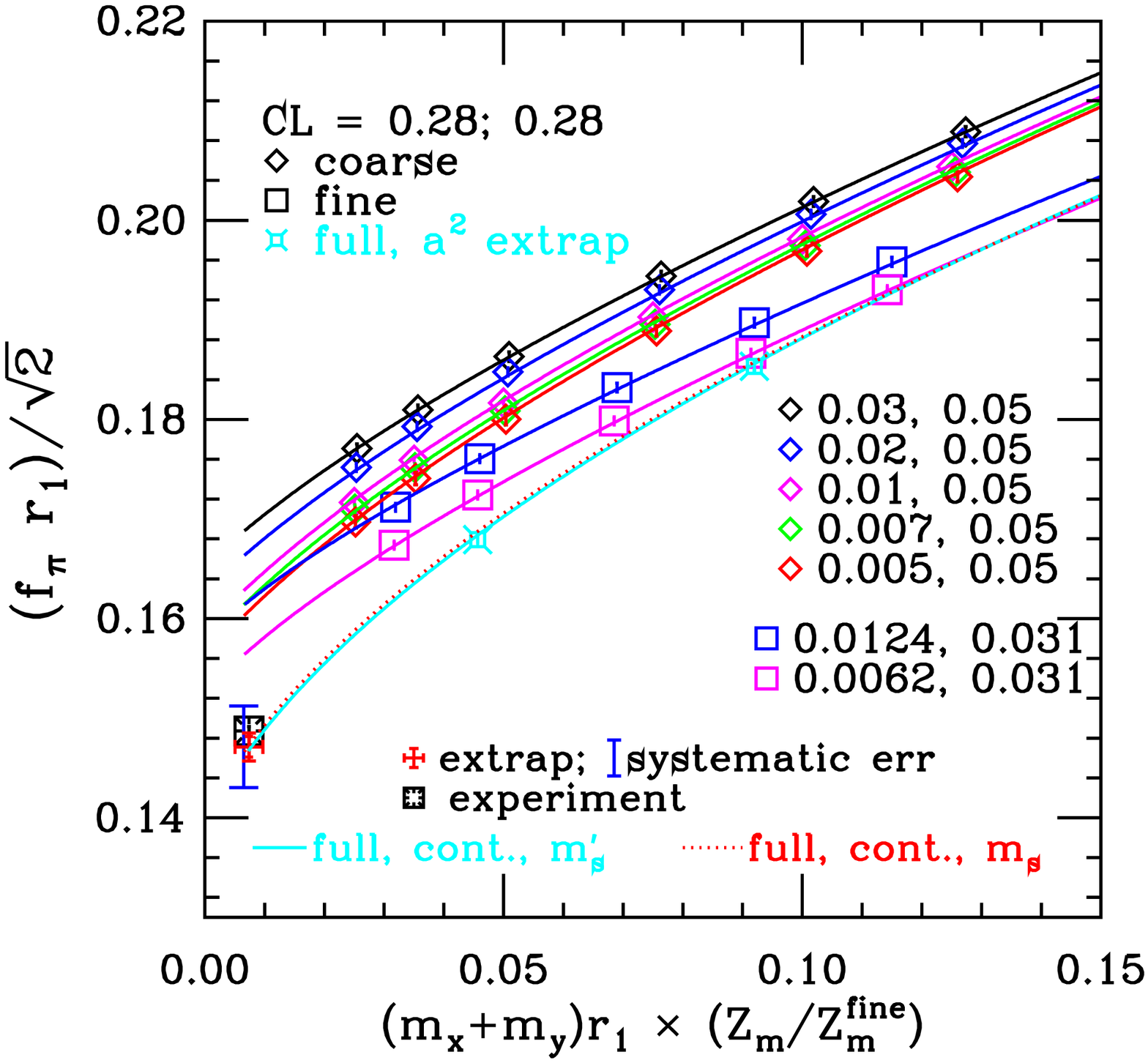}
%\caption{\label{label1}Figure caption for first of two sided figures.}
\end{minipage}\hspace{2pc}%
\begin{minipage}{18pc}
\includegraphics[width=18pc]{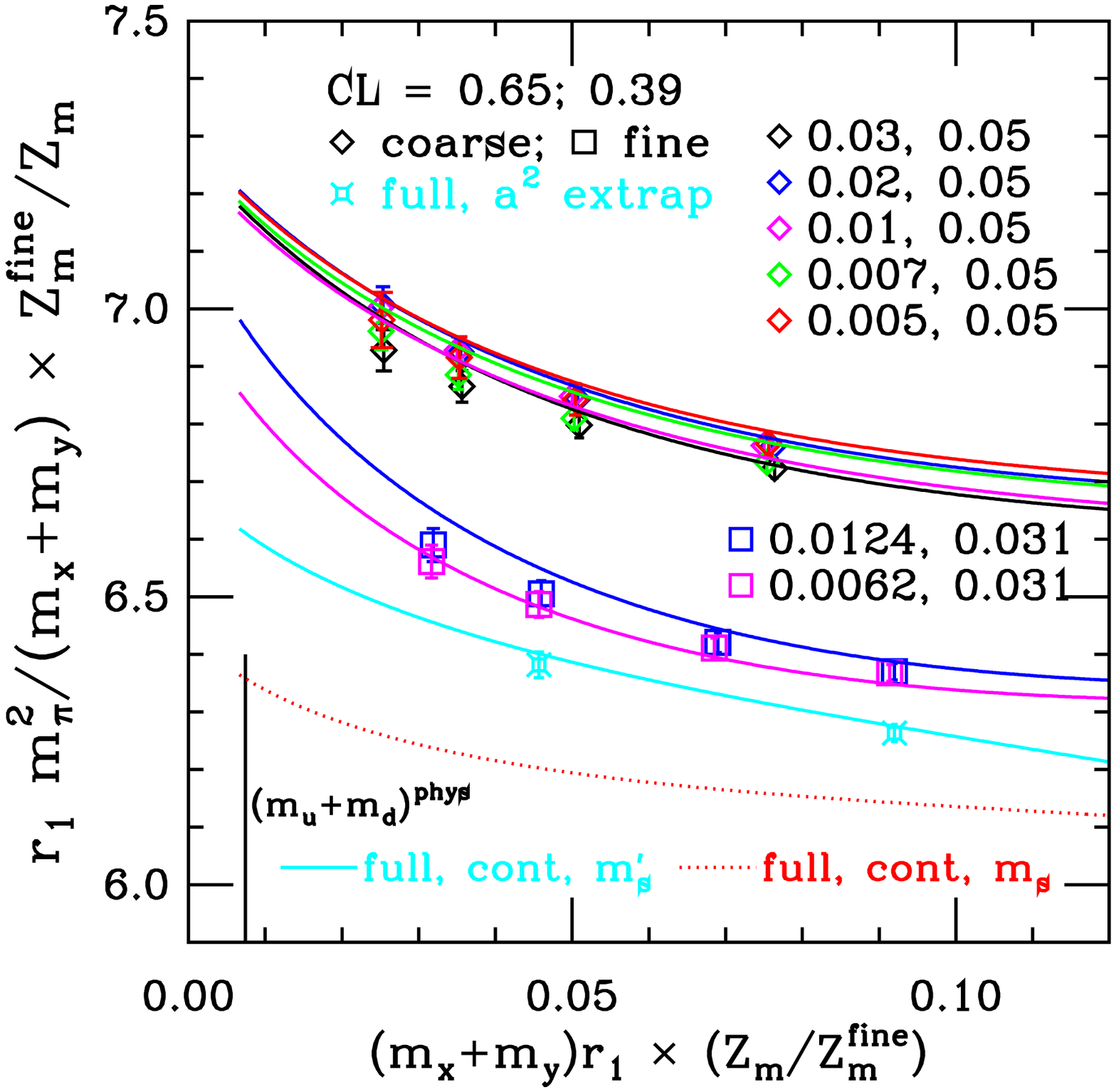}
%\caption{\label{label2}Figure caption for second of two sided figures.}
\end{minipage} 
\caption{\label{chpt_fit}Illustration of \schpt\ fit showing full QCD
points for $f_\pi$ (left) and $m^2_\pi/(m_x+m_y)$ (right). Also shown are
the fits extrapolated to the continuum limit, and after correcting from
the simulation strange quark mass to its physical value.}
\end{figure}

The staggered \chpt\ (\schpt) formalism for our case of 2+1 sea quark
flavors with partially quenched measurements has been worked out in detail
in \cite{CB_SCPT}. All our data for $f_\pi$ and $m_\pi^2$ are fit to the
\schpt\ form simultaneously, taking account of the taste violations.  The
continuum \chpt\ parameters were allowed to have ${\cal O}(\alpha_s a^2)$
terms. After the fit, all the terms taking account of the finite lattice
spacing effects were set to zero to take the continuum limit. Small finite
volume errors ($< 1.5$\%) were corrected within the \chpt\ framework.
Details can be found in \cite{PSEUDO}.  Sample plots for ``full QCD'',
{\it i.e.} with $m_{val} = m_{sea}$ are shown in Fig.~\ref{chpt_fit}.

The strange quark mass in the simulations, our best guess beforehand and
denoted by $m_s'$ in Fig.~\ref{chpt_fit}, turned out a little larger than
the physical strange quark mass, $m_s$. We can correct for this within the
chiral fits. The corresponding curves are also shown in Fig.~\ref{chpt_fit}.

To find the quark masses, we must extrapolate to the physical meson masses.
Electromagnetic and isospin-violating effects ($m_u \ne m_d$) are important.
Denoting by $m_{\hat \pi}$ and $m_{\hat K}$ the masses with EM effects
turned off and $m_u = m_d = \hat m$, {\it i.e.} what is done in our
simulations, we have
\begin{eqnarray}
\label{exp_masses}
m^2_{\hat \pi} & \approx & (m_{\pi^0}^{\rm QCD})^2 \hspace{0.4truecm}
  \approx\hspace{0.4truecm}  (m_{\pi^0}^{\rm expt})^2 \nonumber \\
m^2_{\hat K} & \approx & \frac{(m_{K^0}^{\rm QCD})^2 +
  (m_{K^+}^{\rm QCD})^2}{2}\nonumber \\
(m_{K^0}^{\rm QCD})^2 & \approx & (m_{K^0}^{\rm expt})^2 \\
(m_{K^+}^{\rm QCD})^2 & \approx & (m_{K^+}^{\rm expt})^2
  -(1+\Delta_E)\left((m_{\pi^+}^{\rm expt})^2
  - (m_{\pi^0}^{\rm expt})^2\right)\nonumber
\end{eqnarray}
where the superscript QCD indicates the masses with EM effects turned off.
$\Delta_E=0$ is ``Dashen's theorem.'' Continuum considerations suggest that
$\Delta_E \approx 1$.

With the considerations in eq.~(\ref{exp_masses}) we found, in
collaboration with the HPQCD and UKQCD groups, using a one-loop mass
renormalization constant,~\cite{QMASS}
\begin{eqnarray}
m_s^{\overline{\rm MS}}(2 \; \GeV) &=&  76(0)(3)(7)(0)\;\MeV\ ,\nonumber \\
\hat m^{\overline{\rm MS}}(2 \; \GeV)  &=&   2.8(0)(1)(3)(0)\; \MeV\ , \\
m_s/\hat m  &=&  27.4(1)(4)(0)(1) \ . \nonumber
\end{eqnarray}
Here, the errors are from statistics, simulation, perturbation theory,
and EM effects (obtained from varying $\Delta_E$ between 0 and 2),
respectively.

For the pseudoscalar decay constants we obtain
\begin{eqnarray}
f_\pi & = &  129.5 \pm 0.9\pm 3.5 \; \MeV\ , \nonumber\\
f_K & = &  156.6 \pm 1.0\pm 3.6 \; \MeV\ , \\
f_K/f_\pi  & = & 1.210(4)(13)\ , \nonumber
\end{eqnarray}
where the first error is statistical and the second is systematic.
More details, including our results for the Gasser--Leutwyler chiral
parameters and the ratio $m_u/m_d$ can be found in \cite{PSEUDO}.

Marciano has pointed out~\cite{MARCIANO} that an accurate determination of
$f_K/f_\pi$ can be used to calculate the CKM matrix element $V_{us}$. With
the above result we obtain $V_{us} = 0.2219(26)$. The error, already
comparable to that from the standard method using $K_{\ell3}$ 
decays, is dominated by the lattice error on $f_K/f_\pi$.
With the coming simulations, $V_{us}$ will be known more precisely.

\section{Light hadron spectroscopy}

\begin{figure}[h]
\begin{minipage}{18pc}
\includegraphics[width=18pc]{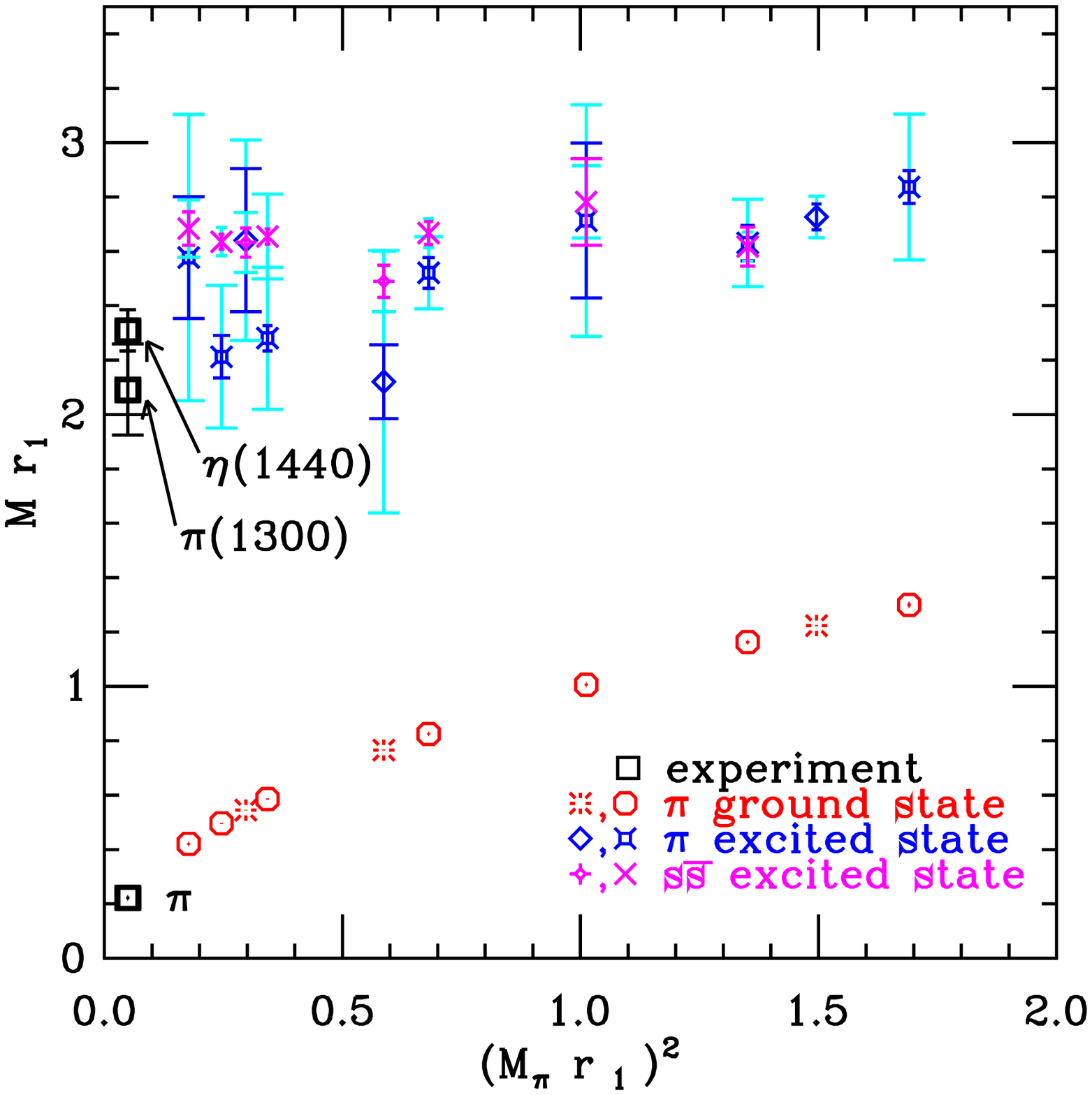}
\end{minipage}\hspace{2pc}%
\begin{minipage}{18pc}
\includegraphics[width=18pc]{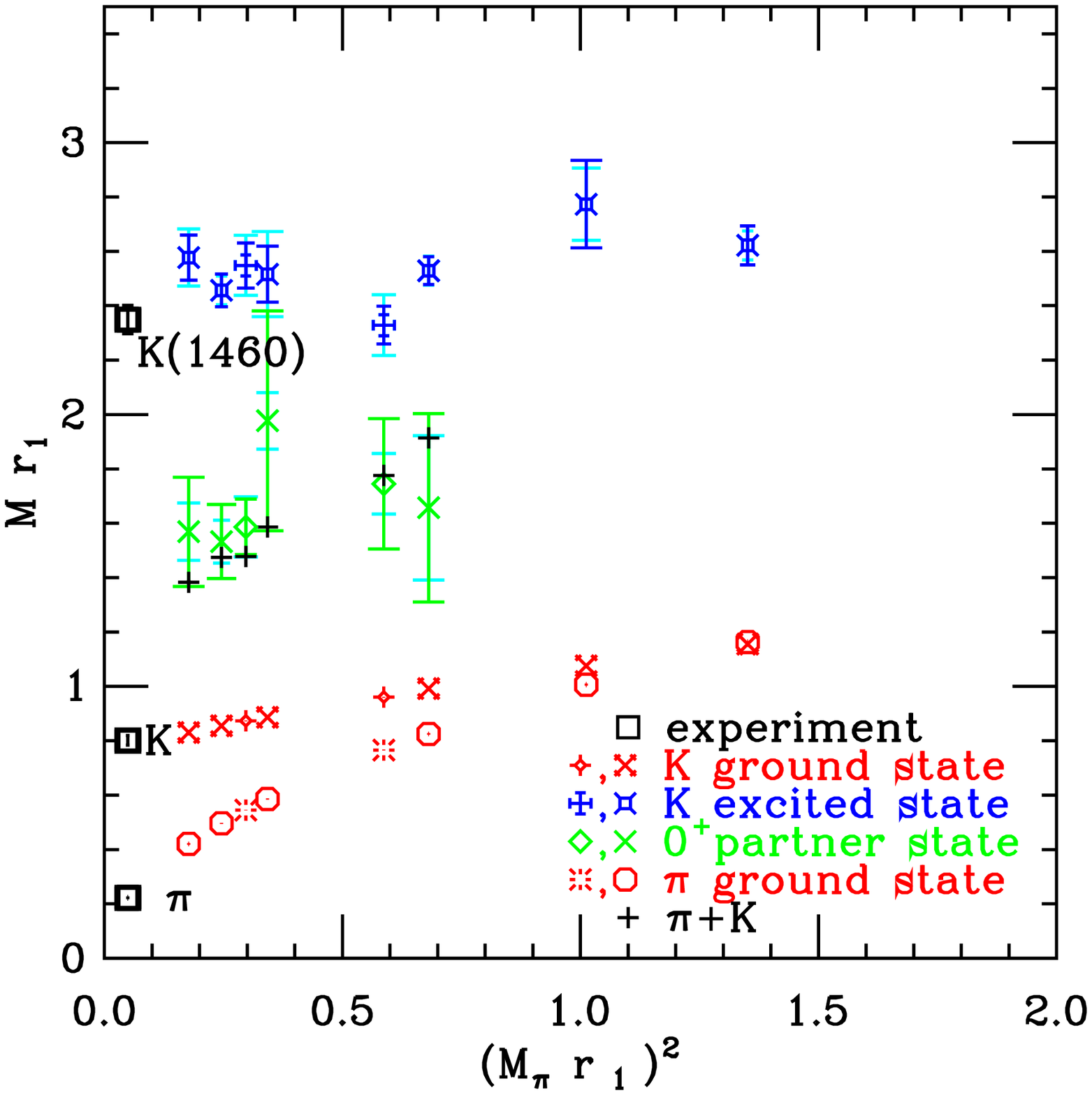}
\end{minipage} 
\caption{\label{excited_PS}Masses of the groundstate and first excited pseudoscalar mesons for pion like states (left) and kaon like states (right).}
\end{figure}

In the pseudoscalar sector the statistical errors on the correlation
functions were small enough that we could get some signal for excited
states, as shown in Fig.~\ref{excited_PS}. It is noteworthy that in kaon
correlation functions we could observe an opposite parity state --
correlation functions with staggered fermions generically contain opposite
parity states -- with energy close to $m_\pi + m_K$, {\it i.e.} the
expected decay products of the opposite parity excited meson. No such
state was found in matched quenched simulations -- absence of virtual
quark loops in quenched QCD prevents emergence of the ``$\pi$ + K''
intermediate state.

\begin{figure}[h]
\begin{minipage}{18pc}
\includegraphics[width=18pc]{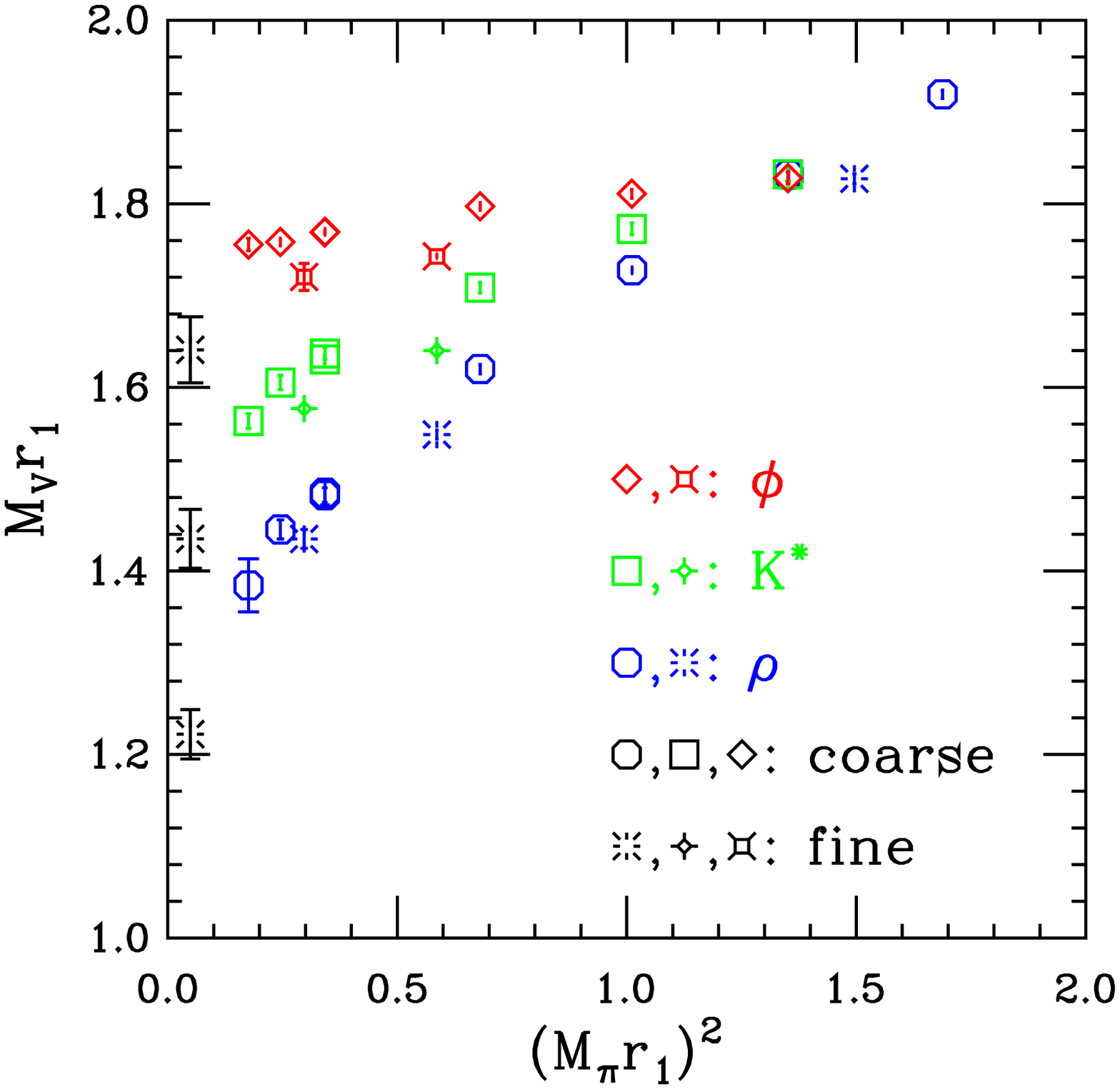}
\caption{\label{vec_mes}Vector meson masses.}
\end{minipage}\hspace{2pc}%
\begin{minipage}{18pc}
\includegraphics[width=18pc]{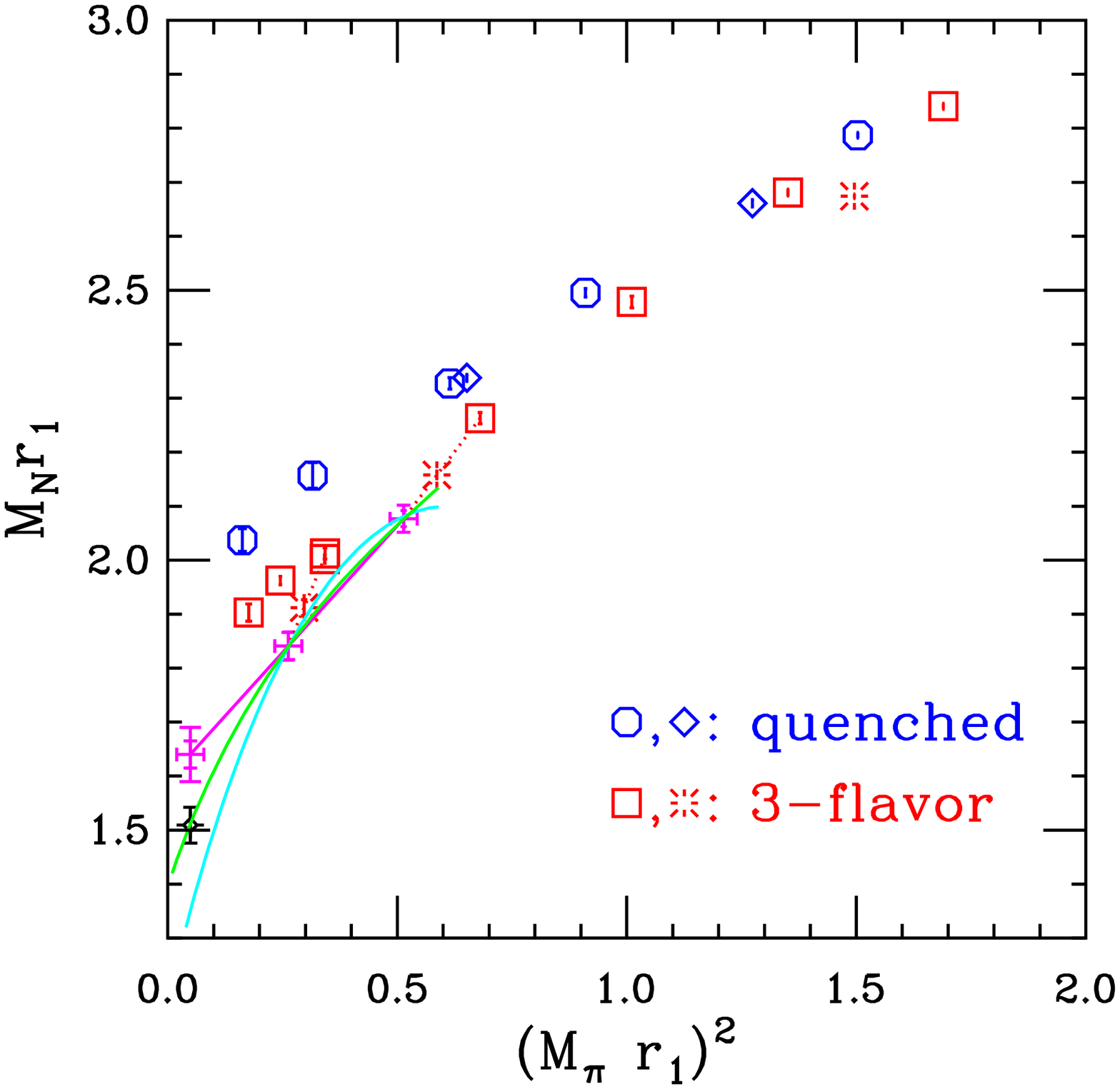}
\caption{\label{baryons}Nucleon masses.}
\end{minipage} 
\end{figure}

Our vector meson and nucleon masses are shown in Figs.~\ref{vec_mes} and
\ref{baryons}. In the latter we also show some possible chiral
extrapolations, after taking the continuum limit at fixed $m_l/m_s$.
For further details and more spectroscopy results see \cite{SPEC}.

\begin{figure}[h]
\includegraphics[width=18pc]{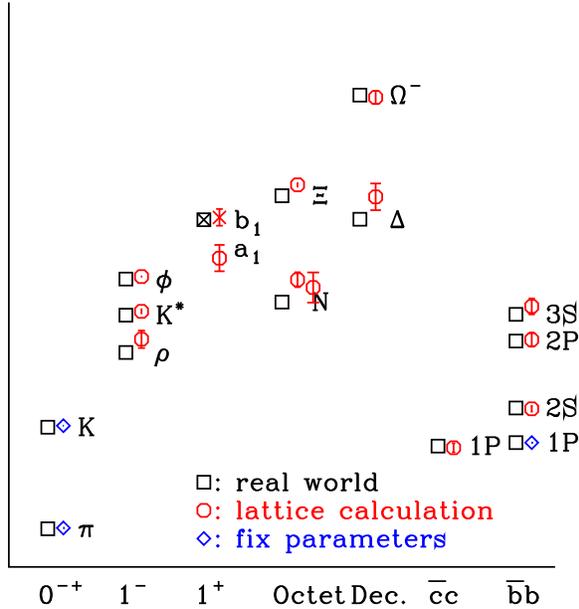}\hspace{2pc}%
\begin{minipage}[b]{18pc}\caption{\label{Bigpic} Comparison of our hadron
spectroscopy, from a crude chiral and continuum extrapolation, with
experiment. $\pi$ and K were used to set the light and strange quark
masses.  The upsilon and charmonium columns are differences from the
ground state (1S) masses, from work of the HPQCD and Fermilab
groups~\cite{PRL,B_s_Wingate}. The $\Upsilon$ 1P-1S splitting was used to
fix the lattice spacing.}
\end{minipage}
\end{figure}

We end by showing a summary of our current spectroscopy results in
Fig.~\ref{Bigpic}.

%\ack
%Simulations were carried out at ....

\end{document}